\documentclass[preprint,numbers,amsmath]{revtex4}
\usepackage{graphicx}
\usepackage{dcolumn}
\usepackage{bm}
\usepackage{epsfig}

\usepackage{wrapfig}

\begin{document}

\title{Momentum space anisotropy of electronic correlations in Fe and Ni - an 
analysis of magnetic Compton profiles}    

\author{L. Chioncel$^{a,b}$, D. Benea$^{c,a}$, H. Ebert$^{d}$, I. Di Marco$^e$ and J. Min\'ar$^{d,f}$}
\affiliation{$^{a}$ Theoretical Physics III, Center for Electronic
Correlations and Magnetism, Institute of Physics, University of
Augsburg, D-86135 Augsburg, Germany}
 \affiliation{$^b$ Augsburg Center for Innovative Technologies, University of Augsburg, 
D-86135 Augsburg, Germany}
\affiliation{$^{c}$Faculty of Physics, Babes-Bolyai University,
Kogalniceanustr 1, Ro-400084 Cluj-Napoca, Romania} 
\affiliation{$^{d}$Chemistry Department, Munich University,
Butenandstr.~5-13, D-81377 M\"unchen, Germany} 
\affiliation{$^{e}$Department of Physics and Astronomy, Division of Materials Theory, 
Uppsala University, Box 516, SE-75120 Uppsala, Sweden}
\affiliation{$^{f}$New Technologies - Research Center, University of West Bohemia,
Univerzitni 8, 306 14 Pilsen, Czech Republic}

\begin{abstract}
The total and magnetically resolved Compton profiles are analyzed within the 
combined density functional and dynamical mean field theory for the transition 
metal elements Fe and Ni. A rather good agreement between the measured and 
computed magnetic Compton profiles (MCPs) of Fe and Ni is obtained with the 
standard Local Spin Density Approximation (LSDA). By including local but 
dynamic many-body correlations captured by Dynamical Mean Field Theory (DMFT), the
calculated magnetic Compton profile is further improved when compared with 
experiment. The second moment of the difference of the total Compton profiles
between LSDA and DMFT, along the same momentum direction, has been used to 
discuss the strength of electronic correlations in Fe and Ni.  
\end{abstract}

\maketitle

\section{Introduction}
The single particle momentum density of an interacting electronic system can be measured 
rather directly by high energy Compton scattering experiments \cite{C85}. These experiments 
in metals provide direct information about 
the occupied momentum states and the Fermi surface. Although the momentum density is a 
relatively simple function it incorporates in a non-trivial way the many-body aspects of the 
interactions between the electrons of the system. 

For several transition metal elements discrepancies between measured and computed Compton profiles 
are found in the low momentum region (Fe, Ni, V, Cr)
\cite{WK77,BZK00,TM06,K04,RWT+73,BMC+12}.
The Compton profile represents a directional property of the investigated system, therefore 
measurements with $p_z$ aligned with various crystallographic directions ({\bf K}) provide 
information related to their structure and the Fermi surface 
through the Compton profile anisotropy. Although the computed anisotropy or difference 
profile, i.e. the difference between two Compton profiles taken along different directions 
(for example {\bf K}=[110] and {\bf K}$^\prime$=[100])  
\begin{equation}\label{Jstruc}
\Delta J^{struc}(p)=J_{110}(p)-J_{100}(p)
\end{equation} 
has in general a trend similar to the experimental spectra, it often displays larger 
amplitudes of oscillations in 
comparison with the measured profiles. The amplitudes of the characteristic oscillations 
are determined by details of the fine structures of the momentum densities. 
Therefore, the {\it structural anisotropies} expressed by Eq. (\ref{Jstruc}) are related 
to some specific features of the Fermi surface topology. In order to address these 
discrepancies, Lam and Platzmann~\cite{LP74a,LP74b} introduced a correction related 
to the difference between the occupation function for a non-interacting electron gas $n^{free}({\bf k})$ 
and a homogeneous interacting electron gas $n^{int}({\bf k})$. This correction takes the form of: 
\begin{equation}\label{JLP}
\Delta J^{LP}(p)=\int \rho({\bf r}) (J^{int}(p)[\rho]-J^{free}(p)[\rho] ) d^3 {\bf r}
\end{equation}
The Lam-Platzman correction Eq.\ (\ref{JLP}) acts in the low-momentum region and for some
cases it reduces the differences between experiment and theory. Nevertheless, the theoretical
values still overestimate the amplitude with respect to the experiment in the low momentum 
region and in addition the residual differences appear anisotropic, contradicting the 
isotropic correction of Lam-Platzmann.
Later on it was suggested by Bauer \cite{BS83,B84} that inclusion of electron-electron 
correlation effects may improve the theoretical difference profiles with 
respect to the experimental measurements. 
The anisotropic effects were modeled for V and Cr by introducing an energy 
dependent occupation function for the d-orbitals \cite{WM90}. While such corrections brought 
the theoretical profile in better agreement with the experiment, one has to stress that this 
has been achieved by incorporating 
the corrections empirically into the calculations. Obviously, the occupation number density in the 
presence of the electronic correlations is non-unity below the Fermi level, the step at $E_F$ is reduced
and becomes non-zero above $E_F$. Kubo~\cite{K01} computed the occupation number
density within the GW approximation 
and discussed the corrections to the Compton profile for the principal directions, concluding that the
strong directional differences are due to the d-bands. 

In this paper we analyze the Magnetic Compton profiles obtained using the combined Density 
Functional and Dynamical Mean Field Theory approach for Ni and Fe. We
supplement our previous results for Ni and Fe \cite{BMC+12} by discussing 
magnetic Compton (MCP) profiles along the $[110]$ and $[001]$ directions. 
The comparison with the experimental data leads us to conclude that theoretical MCP spectra
are improved when local correlations are taken into account. We compute also the total 
Compton (CP) profiles for the main three direction within the cubic symmetry ( $[001]$, $[110]$,  
and $[111]$ directions) at the LSDA and DMFT level. In addition we evaluate the second order moments of 
the difference of Compton profiles taken along the same momentum space direction with and without
including electronic correlations: 
\begin{equation}\label{mcs_dmft-lsda}
\Delta J_{\bf K}(p)= J_{\bf K}^{DMFT}(p) - J_{\bf K}^{LSDA}(p).
\end{equation}
This quantity is different from the 
{\it structural anisotropy} and its second moments $\int_0^{\infty} p^2 \Delta J_{\bf K}(p)$ allows us to discuss the 
{\it momentum space anisotropy of correlations effects} in Fe and Ni.

\section{Computational method}\label{sec:tf}
The electronic structure calculations based on the Density Functional Theory approach 
were performed using the spin-polarized relativistic Korringa-Kohn-Rostoker (SPR-KKR) 
method in the atomic sphere approximation (ASA) \cite{EKM11}. The exchange-correlation 
potentials parametrized by Vosko, Wilk and Nusair \cite{VWN80} were used for the LSDA 
calculations. For integration over the Brillouin zone the special points method has been 
used \cite{MP76}. In addition to the LSDA calculations, a charge and self-energy 
self-consistent LSDA+DMFT scheme for correlated systems based on the KKR
approach~\cite{MCP+05,MM09,MI11} has been used. 
The many-body effects are described by means of dynamical mean field theory (DMFT)~\cite{MV89,GK96,KV04}
and the relativistic version of the so-called Spin-Polarized T-Matrix Fluctuation Exchange 
approximation \cite{KL02, PKL05} impurity solver was used. The realistic multi-orbital 
interaction has been parametrized by the average screened Coulomb interaction $U$ and 
the Hund exchange interaction $J$. The values of $U$ and $J$ are sometimes used as fitting 
parameters, although recent developments made it in principle possible to compute the 
dynamic electron-electron interaction matrix elements with a good accuracy \cite{AIG+04}. The
static limit of the screened energy dependent Coulomb interaction leads to a $U$ parameter 
in the energy range between 2 and 4 eV for all 3d transition metals, with
substantial variations associated to the choice of the local orbitals \cite{MA08}.
As the $J$ parameter is not affected by screening it can be calculated directly within the LSDA and
is approximately the same for all 3d elements, i.e $J$ $\approx$ 0.9 eV. In our calculations we 
used values for the Coulomb parameter in the range of $U$ = 2.0 to 3.0 eV and the Hund 
exchange-interaction $J$ = 0.9 eV.

The KKR Green function formalism was recently extended to compute magnetic Compton profiles 
(MCPs)~\cite{SGST84,BME06,DB04}. In the case of a magnetic sample the spin resolved momentum 
densities are computed from the corresponding LSDA(+DMFT) Green functions in momentum space as:
\begin{equation}\label{e7}
n_{m_s}(\vec p)={-\frac{1}{\pi} \Im    \int_{-\infty}^{E_F}
G_{m_s}^{LDA(+DMFT)}(\vec p,\vec p,E)dE} \; ,
\end{equation}
where $m_s=\uparrow(\downarrow)$.

The momentum density defined as $n_{\uparrow}(\vec p) + n_{\downarrow}(\vec p)$
projected onto the direction ${\bf K}$ defined by the scattering vector, allows to define 
the Compton profile as a double integral in the momentum 
plane perpendicular to the scattering momentum $\vec p_z$:
\begin{equation}\label{mcs3}
J_{\bf K}^{LDA(+DMFT)}(p_z)=
\int \int [ n_{\uparrow}(\vec p) + n_{\downarrow}(\vec p) ] dp_x dp_y; (p_z || {\bf K}).
\end{equation}
Analogously, the double integral of the spin momentum density $n_{\uparrow}(\vec p) -
n_{\downarrow}(\vec p)$ projected onto the scattering direction defined by the
vector  ${\bf K}$ defines the magnetic Compton profile (MCP):
\begin{equation}\label{mcsMCP}
J_{mag,\bf K}^{LDA(+DMFT)}(p_z)=
\int \int [ n_{\uparrow}(\vec p) - n_{\downarrow}(\vec p) ] dp_x dp_y; (p_z || {\bf K}).
\end{equation}
  
The electron momentum densities are usually calculated for the
principal directions ${\bf K}=[001], [110], [111]$ using an rectangular grid
of 200 points in each direction. The maximum value of the
momentum in each direction is 8 a.u.. The Magnetic-Compton/Compton
profile is normalized such that the area under its curve is equal
to the magnetic moment / number of valence electrons. This means that for the
ordinary Compton profile the contribution of the core electrons has been
omitted, as this does not show an anisotropy.

\section{Electronic correlations in Fe and Ni}
\label{sec:corel_fe_ni}
It is commonly accepted that the decisive features of ferromagnetic Fe and Ni are determined 
by the electronic correlation effects taking place in the relatively narrow $3d$ band, 
which hybridizes weakly with the $4s$ and $4p$ bands. Fe (Ni) has a cubic body (face)-centered 
structure with lattice parameter 2.86 (3.52) $\AA$ \cite{LC91} and  8 (10) electrons within the
valence band, about 7 (9) of them having predominantly d-character.
Important differences between Ni and Fe are the following: Ni has a rather small exchange 
splitting of about $0.2-0.3$ eV \cite{EH78,DG78,HK79,EH80,EP80}, while in Fe this is more 
substantial and amounts to
$2.2-2.4$ eV \cite{KS85,KG84}, i.e. a difference by a factor ten. Ni exhibits a 
prominent satellite structure at about $6$ eV below the chemical potential \cite{GBP+77}, while
the existence of an analogous feature in Fe is still controversial~\cite{GM07}. 
On the other hand, Fe exhibits an ``exchange splitting'' persisting into the
high temperature phase, while in Ni such a feature seems absent. 

From a theoretical point of view, band structure calculations based on DFT are able 
to account for ground state properties of Fe quite reasonably. Even the most striking failure 
of LSDA, namely the prediction of an fcc instead of the experimental bcc ground state in Fe, 
is explained by the tiny energy difference between the two structures within GGA
\cite{BJ89,BM90,LC90,SP91}. 

State-of-the art 
computations including many-body effects were recently used to scrutinize the paramagnetic
 $\alpha$-phase of iron. An orbital selective local moment formation mechanism was proposed
\cite{KP10}. Later on Leonov et al. introduced the correlation magnetic energy and for
the first time explained the $\alpha$-to-$\gamma$ phase transition in paramagnetic iron \cite{LP11}.
Subsequently this opened the path towards the computation of the phonon spectra across 
the $\alpha$-to-$\gamma$ phase transition and the study of lattice stability in the 
presence of electronic correlations~\cite{LP12}. Concerning the methodological
background, the generalization to a rotational invariant exchange interaction 
allowed to revisit the magnetic properties of paramagnetic $\alpha$ iron~\cite{AB12} 
and to establish a reasonably good agreement for the Curie temperature of 
Fe and Ni \cite{BL13}.
A remarkable difference between Fe on the one side and Ni on the other side lies in the fact
that in the latter the majority spin bands are fully occupied, while this is not the
case in Fe.
The LSDA calculations for fcc Ni cannot reproduce some features of the electronic structure 
of Ni as observed experimentally. The valence band photoemission spectra of Ni shows a 3d-band 
width that is about $30 \%$ narrower than obtained from the LSDA
calculations. It is known from VB-XPS spectra 
that LSDA cannot reproduce the dispersionless feature at about 6 eV binding energy (the 
so-called 6 eV satellite). In addition the magnetic exchange-splitting is overestimated by 
LSDA calculations when compared with the experimental data. An improved description of 
correlation effects for the 3d electrons via LSDA+DMFT gives a more correct width 
of the occupied 3d bands, a better exchange splitting, and also the 6 eV satellite structure 
in the valence band \cite{LK01,CVA+03,MCP+05,BME+06,GM07,SFB+09,GM12,SBM+12}.

Concerning the magnetic Compton profiles of Fe and Ni, the experimental
spectra and the FLAPW calculations
based on LSDA are in fair agreement \cite{KA90}. For Fe the center of gravity 
of p-states were lowered to reproduce correctly the N-centered hole pocket of the third 
minority-spin band \cite{KA90}. This shows that LSDA needs to be supplemented to
obtain a better description of the MCP. For Ni a slightly noticeable discrepancy in 
the spectra are seen. 
In the literature discrepancies between calculated and experimental MCPs are often
 attributed to non-local corrections to
 the potential stemming from electronic correlations. However, in order to
 check which prescription beyond 
the LSDA potential performs better, we first take into account local dynamic electronic 
correlations. Clearly, on the other hand,  measurements with higher
statistical accuracy are also desired, in order to provide a critical test of band theories.

\subsection{Magnetic Compton profiles of Iron} %[110] and [100]} 

The magnetic Compton profiles along the $[111]$ direction for Fe and Ni 
including dynamic correlations were studied recently by Benea et al.~\cite{BMC+12}. 
Here we extend this study including results for the $[001]$ 
and $[110]$ directions for both magnetic and non-magnetic Compton profiles.

\begin{figure}[h]
%\vspace{0.5cm}
%\begin{center}
%\hspace*{0.2cm}
   \includegraphics[width=0.99\linewidth, clip=true]{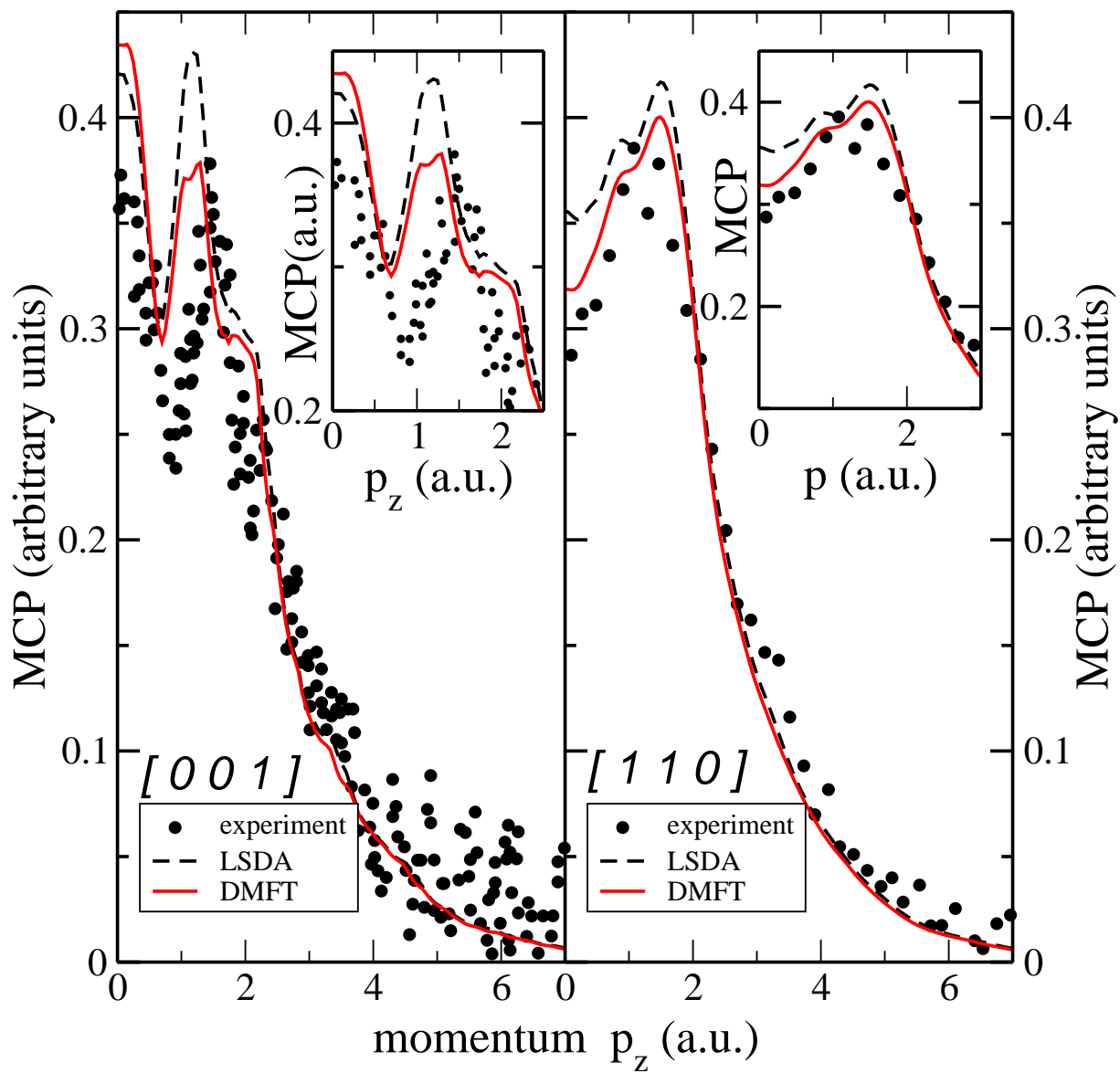}
%\end{center}
\caption {\label{Fig:figure1}(Color on-line) Magnetic Compton profiles of Fe along 
the [001] and [110] directions, calculated with LSDA and LSDA+DMFT with $U$ = 2.3 eV, 
$J$ = 0.9 eV and $T$=400K. The MCP profiles were convoluted with the experimental resolution 
(0.12 a.u. for [001] MCP and 0.7 a.u. for [110], respectively). The experimental 
MCPs stem from Sakurai et al. \cite{STO94} and Collins et al. \cite{CCT+89}. 
}
\end{figure}

The computed magnetic Compton profiles along the [001] and [110] directions of Fe are 
shown in Fig. \ref{Fig:figure1}. The dashed (solid, red) curve 
represents the results of LSDA (LSDA+DMFT) calculations. The average
Coulomb U=2.3 eV and exchange J=0.9 eV parameters have been used, and the 
temperature was taken as 400K.  The experimental MCP data 
are taken from the work of Sakurai et al.~\cite{STO94} for the [001] 
direction while for the [110] direction the results presented in the paper
of  Collins et al.~\cite{CCT+89} are given. The experimental momentum 
resolutions are 0.12 a.u.\ for the [001] and 0.7 a.u.\ for the [110] direction,
respectively. According to this, the theoretical spectra have been 
convoluted with Gaussians corresponding to the experimental resolution. 
After convolution, the calculated MCPs have been scaled to correspond to a spin momentum 
of 2.3 $\mu_B$ (LSDA) and 2.19$\mu_B$ (LSDA+DMFT), respectively.
    
%%%%%%%%%%%%%%%%%%%%%%%%%%%%%%%%%%%%%%%%%%%%%%%%%%%%%%%%%%%%%%%%
%%%%%%%%%%%%%%%%%%%%%%%%%%%%%%%%%%%%%%%%%%%

As one can see from Fig. \ref{Fig:figure1} there is a fair agreement between the
measured and computed MCP spectra except for the region with momenta $p_z < 2.5$ a.u.  
where noticeable differences are visible.
The spectra change shape in the small momentum region $p_z <2.5$ a.u., with
a depletion around $1$ a.u. for the [001] direction. On the other hand, for the
[110] direction no such depletion is seen. In the high-momentum
region $p_z> 2$ a.u., structureless similar shapes are observed for both
[001] and [110] directions.
Similarly to the results discussed in our recent work \cite{BMC+12} for the 
[111] direction, we see that DMFT improves the agreement with the experimental 
spectra for the [110] direction at small momenta $ p_z < 1.5$ a.u.. 
A different situation is noticeable for the spectra along [001]: in the vicinity
of zero momentum DMFT results slightly overestimates experiment and follow
very closely the LSDA data until $ p_z = 1.5$ a.u.. The maxima in the LSDA profile
at $\approx 1.25$ a.u. is significantly reduced. However, this reduction is not sufficient 
to intercept the experimental data. At higher momenta both LSDA and DMFT lead
to similar MCPs. Although the experimental momentum resolution is 
rather satisfactory, at large momenta the spread in the experimental data,
in particular along $[001]$ indicates the need for enhanced accuracy in the experiment.

\begin{figure}[h]
%\vspace{0.5cm}
%\begin{center}
%\hspace*{0.2cm}
   \includegraphics[width=0.99\linewidth, clip=true]{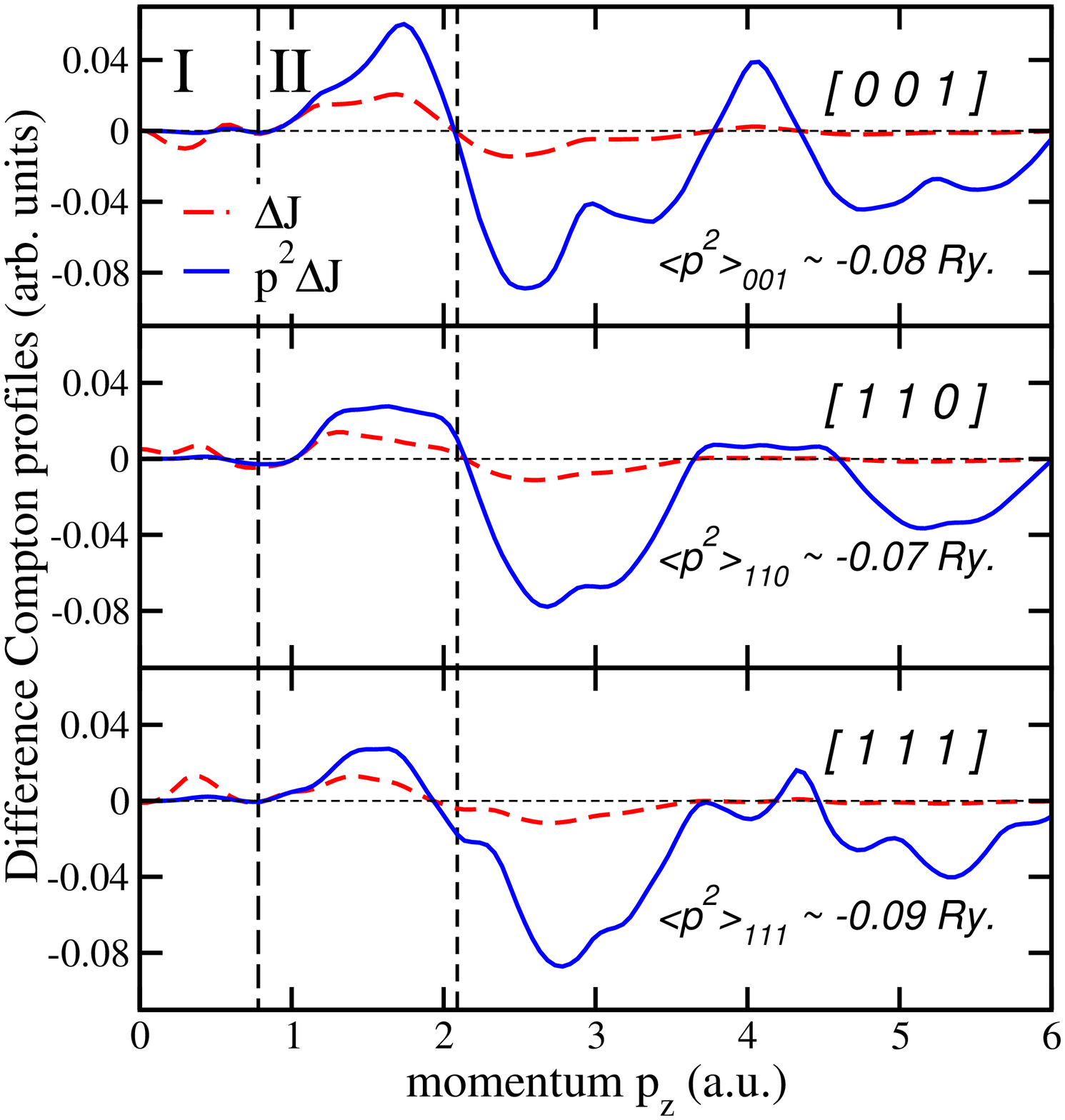}
%\end{center}
\caption {\label{Fig:figure2}(Color on-line) Differences between the LSDA+DMFT
and LSDA total Compton profiles of Fe $\Delta J$  (red dashed) and its second
order moment $p^2 \Delta J$ (blue solid) along [001], [110] and [111] directions.
Larger values of the second moment indicate a more important electronic correlation
energy contribution for the corresponding direction. }
\end{figure}

To discuss further the characteristic features of the correlations we plot in 
Fig.~\ref{Fig:figure2} the difference between the total Compton profiles obtained 
within the LSDA- and DMFT-based calculations together with its second moment
for the main three directions. A similar trend in the momentum dependence is seen.

%%%%%%%%%%%%%%%%%%%%%%%%%%%%%%%%%%%%%%%%%%%%%%%%%%%%%%%%%%%%%%%%
%%%%%%%%%%%%%%%%%%%%%%%%%%%%%%%%%%%%%%%%%%%

Essentially, one can distinguish three regions: the low momentum region (I) 
for momentum smaller than $\approx  0.8$ a.u., the intermediate region (II) 
$0.8$ a.u.$ < p_z < 2$ a.u. and finally the high momentum region  $p_z > 2$ a.u. 
The comparison between different directions shows differences concerning the 
shape as well as the absolute values. The most significant discrepancies are seen at low 
momenta ($p_z \le 1$ a.u.): along the $[001]$ direction, one can 
find regions in which electronic correlations
lead to depleted ($\Delta J <0$) or enhanced ($\Delta J>0$) Compton
profiles. In contrast to the other two $[110]$ and $[111]$ directions with $\Delta J >0 $ the Compton profile is 
enhanced because of Coulomb interactions. The maximum at $p_z=0$ remains also along the [110] 
direction, while for [111], the difference $\Delta J$ shows a peak and a central dip. 
Within the region of the intermediate momenta $1$ a.u. $ < p_z < 2 $ a.u.
$\Delta J$ has positive values along the [001] and [111] directions and small 
negative (around $1$ a.u.) as well as positive values for [110]. Thus, the electronic 
correlations lead to an overall
enhancement of the Compton profile in the intermediate region. 
For all principal directions $\Delta J$ behave similarly in the high momentum region,
being essentially negative in the entire range, with a slightly positive hump at 
$p_z = 4 $ a.u. for the [001] direction.

A more quantitative analysis upon the momentum space anisotropy of correlations 
can be made by calculating the second moment of the Compton profile. The second 
moment has been previously applied to study the redistributions of interatomic
interactions in the momentum densities, which allowed to connect the Compton 
profile with the interaction energy and interatomic forces \cite{KM82,TH83}. 
Taking the second moment along the bond directions allows to study the electronic properties
of the bond in momentum space. In coordinate space the charge is contracted around the 
nucleus and accumulated along the bond direction. The reverse of the situation 
happens in momentum space: momentum density is greater perpendicular to the bond 
direction~\cite{C85}. In the same spirit, 
it is possible to compute the second moment of the difference between 
correlated and non-correlated Compton profiles, along the bond directions ${\bf K}$:
\begin{equation}
\langle p^2 \rangle_{\bf K}=\int_0^{\infty} p_z^2 \left[ J^{DMFT}_{\bf K}(p_z) - 
J^{LSDA}_{\bf K}(p_z) \right] dp_z;   (p_z || {\bf K})
\end{equation}
which allows to discuss the effects of the electronic interactions upon the bounded 
density. 

In the case of Fe the values for the second moment of the difference in the total Compton 
profiles are given in Fig.\ \ref{Fig:figure2}. We observe that including electronic interactions treated
beyond mean-field, the second moment of the difference decreases along all bonds. We have 
obtained a stronger decrease along the [111] and [001] direction, and weaker decrease along 
[110]. We note also that the decrease happens in agreement with the interatomic distances in 
the bcc lattice: for a shorter bond a stronger decrease is evidenced. These results demonstrate 
that (i) although the included interaction is only local, its consequences, i.e. the electronic 
correlations, show momentum space anisotropy; (ii) in addition shorter bonds experience stronger
effects.
A more detailed discussion concerning the connection between the second moment 
along a ${\bf K}$ direction  and the energy of an interacting electronic system is provided 
in section Sec.\ref{second+E}.

\subsection{Magnetic Compton profiles of Nickel} % [110] and [100]} 

\begin{figure}[h]
%\vspace{0.5cm}
%\begin{center}
%\hspace*{0.2cm}
   \includegraphics[width=0.99\linewidth, clip=true]{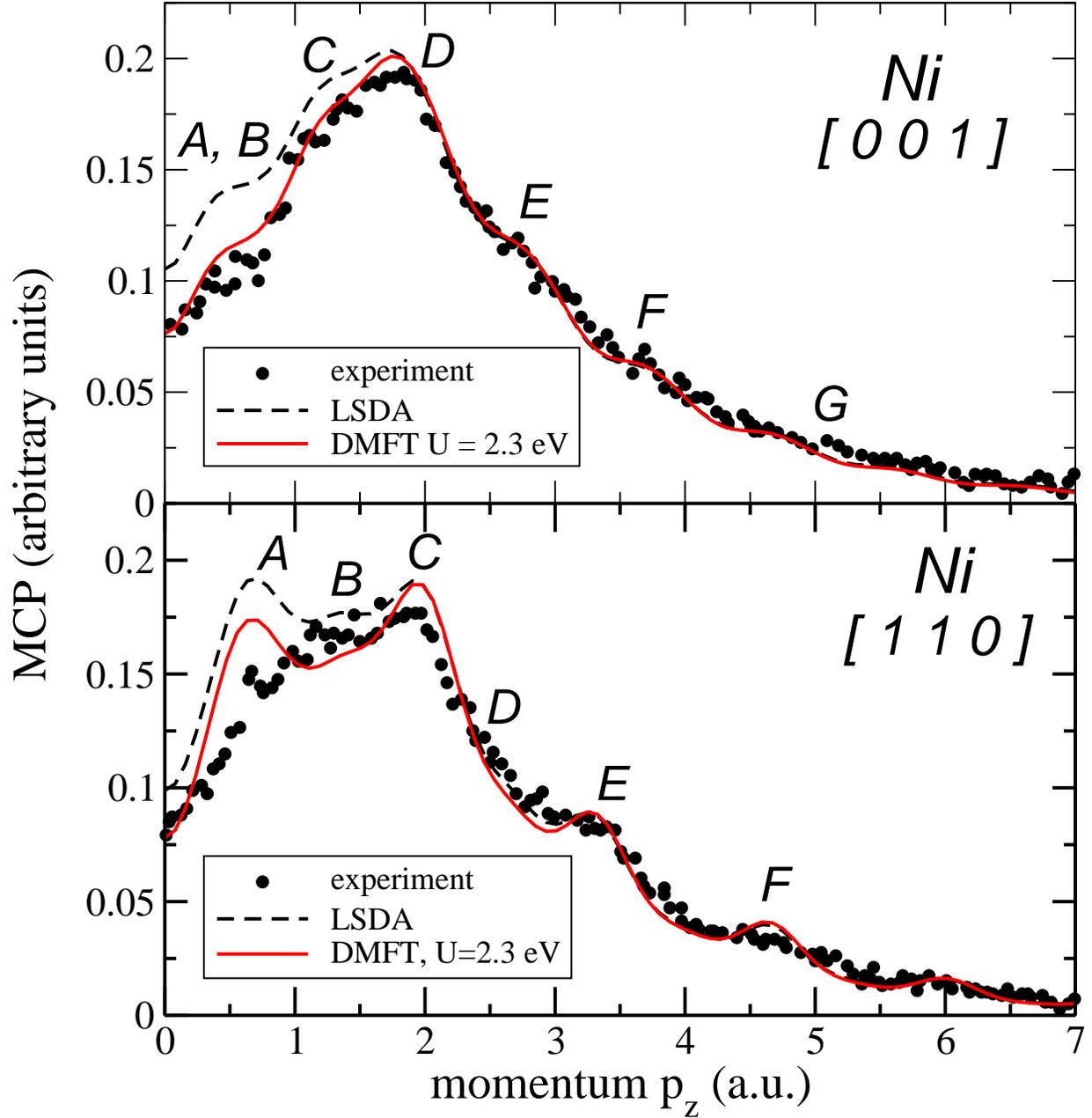}
%\end{center}
\caption {\label{Fig:figure3}(Color on-line) Magnetic Compton profiles of Ni along [001] 
(upper panel) and [110] (lower panel) directions, computed with LSDA and LSDA+DMFT with 
$U$ = 2.3 eV, $J$ = 0.9 eV and $T$=400K. The computed data were convoluted according to the experimental 
resolution of 0.43 a.u.. The experimental profiles are taken from Ref. \cite{DDG+98}.}
\end{figure}

The magnetic Compton profiles of Ni along the [001] and [110] directions
are shown in Fig.\ \ref{Fig:figure3}. The dashed (solid, red) curve 
represents the LSDA (LSDA+DMFT) calculations. The theoretical calculations
are compared with the experimental MCP data of Dixon et al. \cite{DDG+98}.
The experimental momentum resolution is 0.43 a.u., which was also used 
as Gaussian broadening parameter for the calculated MCP spectra. In addition, 
the calculated MCPs have been scaled to the spin magnetic moment
0.6 $ \mu_B$ for LSDA and to 0.6, 0.57, 0.56 and 0.5 $\mu_B$ for 
the corresponding values of U = 1.8, 2.0, 2.3 and 3.0 eV for the LSDA+DMFT 
calculations. 

%%%%%%%%%%%%%%%%%%%%%%%%%%%%%%%%%%%%%%%%%%%%%%%%%%%%%%%%%%%%%%%%
%%%%%%%%%%%%%%%%%%%%%%%%%%%%%%%%%%%%%%%%%%%

As one can see in Fig. \ref{Fig:figure3} (upper panel) LSDA results are already 
in reasonable agreement with the measurements. They capture the behavior at 
large moments, get close to the maximum at $\approx 1.8$ a.u. and overestimate 
the contributions in the low momentum region. Dixon et al. \cite{DDG+98} analyzed 
the magnetic Compton profile of Ni comparing LSDA and GGA results obtained through 
a linear muffin-tin orbitals (LMTO) method. In their theory, and in particular 
from the analysis of the fifth band, Dixon et al. \cite{DDG+98} identified several
main peaks, which they labeled from A until G. In the present study all these major
features are essentially reproduced, although they are not very evident in our plot,
due to the Gaussian broadening. The first two peaks, which LSDA locates at
0.3/0.7 a.u (inside the first Brillouin zone), are not resolved by the experiment.
The highest peaks, labeled as C and D, are located at 1.25/1.70 a.u., and are followed
by other Umklapp peaks at 2.7 (E), 3.6 a.u (F) and a further shoulder at G. 
The LSDA results (dashed black line) overestimates significantly the contributions in 
the low momentum region, while the DMFT profile is in much better agreement with 
experiments, until the peak D. There are no essential differences between the 
LSDA and LSDA+DMFT spectra for momenta larger than $\approx 1.7$ a.u..

The lower panel of Fig. \ref{Fig:figure3}, shows the MCPs along the [110] direction.
As for the [001] direction, our results are in good agreement with previous 
results by Dixon et al. \cite{DDG+98}. Following their notation, a first peak A is 
situated inside the first Brillouin zone, and located 
around 0.7 a.u.. All subsequent peaks are essentially of Umklapp origin, and the 
maximum of the MCP is at C, being overestimated in theory in comparison with
experiment. It was remarked by Dixon et al. \cite{DDG+98} that all computed 
peaks at higher momenta E (3.3 a.u) and F (4.7 a.u.) are more visible than the 
corresponding maxima in the experiment. This seems to hold also for DMFT
results. Instead, the low momentum region ($ p_z < 1.7$ a.u.) is in better
agreement with experiment, in particular for the value at zero momentum 
and for the peak within the first Brillouin zone (A). Further, the experimental 
value of the MCP at B seems to be at an intermediate level between LSDA and DMFT.
Both LSDA  and DMFT overestimate the maximum C from where they follow a very similar 
momentum dependence, as mentioned above.
 
%%%%%%%%%%%%%%%%%%%%%%%%%%%%%%%%%%%%%%%%%%%%%%%%%%%%%%%%%%%%%%%%
\begin{figure}[h]
%\vspace{0.5cm}
%\begin{center}
%\hspace*{0.2cm}
   \includegraphics[width=0.99\linewidth, clip=true]{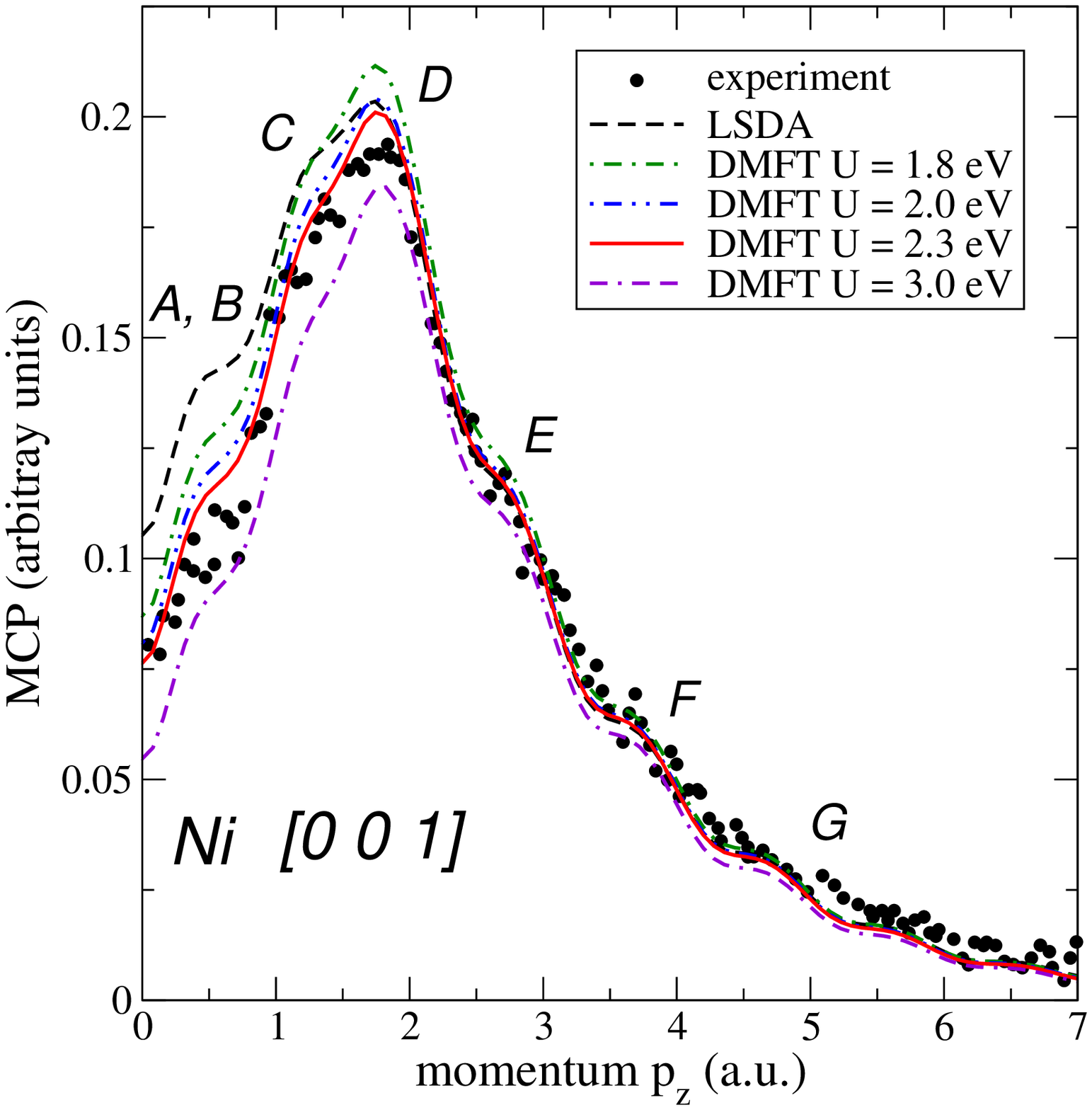}
%\end{center}
\caption {\label{Fig:figure4}(Color on-line) Magnetic Compton profiles of Ni along 
[001] direction. Black dashed-solid LSDA, red solid-line LSDA+DMFT for $U$ = 2.3 eV, 
dot-dashed-blue line $U$ = 2.0 eV and dot-dot-dashed $U$ = 3.0 eV. 
In all calculations $J$ = 0.9 eV and $T$=400K. The 
experimental data is taken from Ref. \cite{DDG+98} (momentum resolution 0.43 a.u.).}
\end{figure}
%%%%%%%%%%%%%%%%%%%%%%%%%%%%%%%%%%%%%%%%%%%%%%%%%%%%%%%%%%%%%%%%

The MCP for Ni along the [001] direction using LSDA and LSDA+DMFT are
presented in Fig.~\ref{Fig:figure4} for different values of the $U$ 
parameter. The experimental data are shown with dots and the LSDA 
results are given by dashed lines. 
The DMFT results are presented for $U$ = 1.8, 2.0, 2.3 and 3.0 eV and are respectively
plotted with green (dot-dashed), blue (dot-dot-dashed), red (solid) and violet
(dashed-dot-dashed) lines. The values in the region of low momentum, up to
$p_z < 1$ a.u. (most of it in the first Brillouin zone), are better captured
by the intermediate values of $U$ = 2.3 eV. A slightly over- and a more significant
underestimation can be noticed for $2.0$ eV and $3.0$ eV, respectively. Around the maximum 
($p_z \approx 1.7$ a.u.) of the experimental profile (label D) LSDA and all  
DMFT results $U$ =1.8, 2.0 and 2.3 eV overestimate the magnitude of the maxima, 
except $U$=3 eV which underestimates the contribution at this position. For momenta larger
than point D the DMFT profiles with $U$ = 2.0 and 2.3 eV and the LSDA profile have 
essentially the same behavior, in good agreement with experimental data. Dixon et
al. \cite{DDG+98} noted that discrepancies may not be eliminated 
simply renormalizing the magnetic moment because the moment is connected to the exchange 
splitting, therefore this would not necessarily scale the MCP spectra. 
Figs. \ref{Fig:figure3} and Fig. \ref{Fig:figure4} show that electronic correlations
beyond LSDA/GGA improve the spectra in the low momentum region.  We mentioned above
(Sec. \ref{sec:corel_fe_ni}) that the reduction of exchange splitting
is one among several subtle consequences of the correlation effects in Ni/Fe. 
Therefore DMFT accounts naturally for the renormalization, in this case the 
reduction of the magnetic moment.  Obviously, this has consequences on the entire
momentum dependence of the MCP spectrum, also in the high momentum region.
The high momentum behavior was attributed to a free-atom type profile 
\cite{CD00}, based on the argument that the strong weighting of the high momentum 
components into the sum of the second moments dominates the cohesive energy. This 
argument may be invoked also in the presence of electronic correlations, although 
the non-interacting and interacting case have different atomic-limits. 
As Figs. \ref{Fig:figure3} and \ref{Fig:figure4} show in the high momentum region 
very similar LSDA and DMFT profiles, electronic correlations seem to have little
influence in this region. To learn more about the consequence of electronic 
correlations in the following we analyze the difference of the total Compton 
profiles with and without correlations and its second order moments.
 
%%%%%%%%%%%%%%%%%%%%%%%%%%%%%%%%%%%%%%%%%%%%%%%%%%%%%%%%%%%%%%%%
\begin{figure}[h]
%\vspace{0.5cm}
%\begin{center}
%\hspace*{0.2cm}
   \includegraphics[width=0.99\linewidth, clip=true]{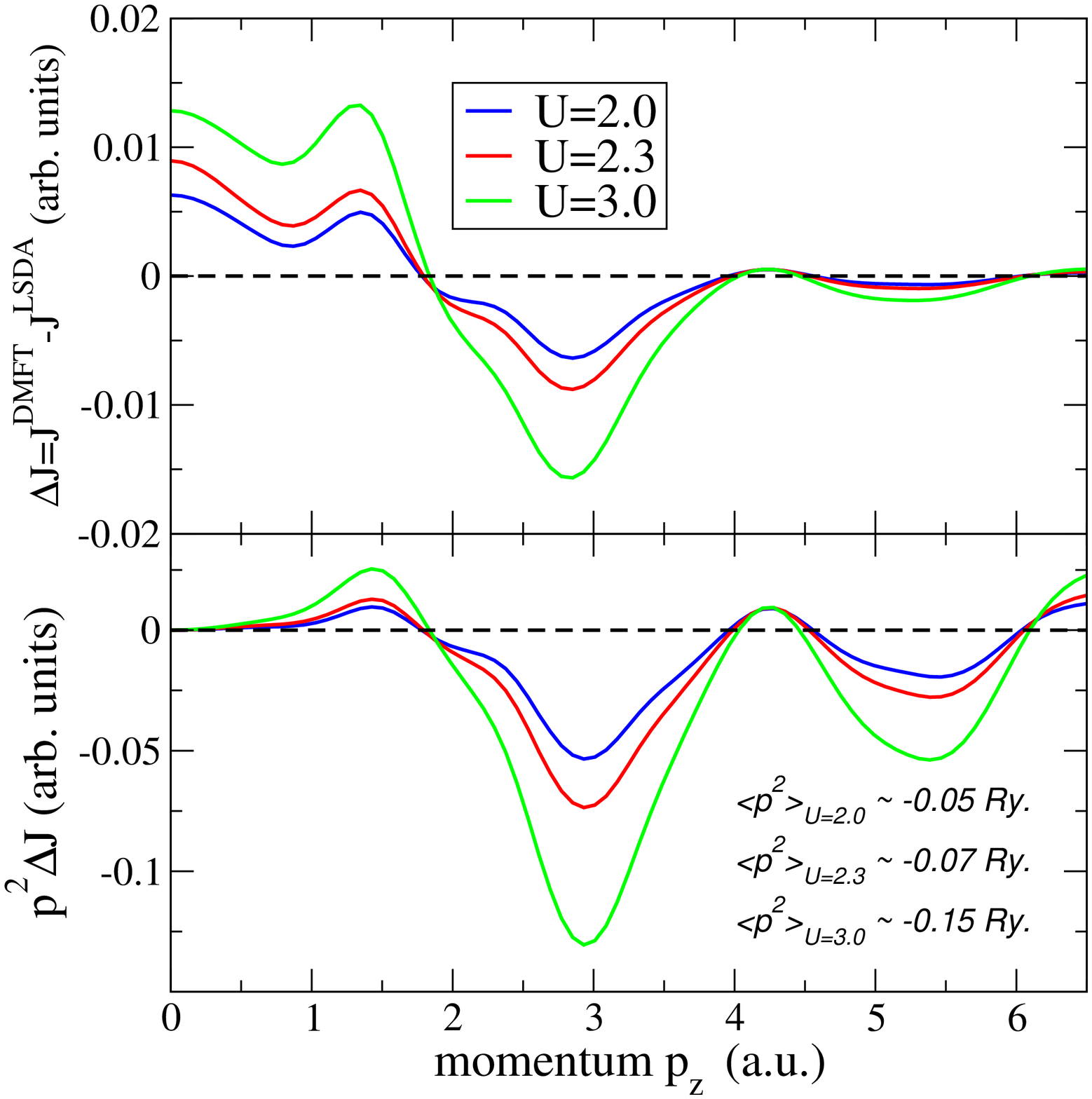}
%\end{center}
\caption {\label{Fig:figure5}(Color on-line) Difference of the Compton
profile (top) and $p^2 \Delta J$ (bottom), together with the second moments.
Values for the second moments are given in Ry.}
\end{figure}
%%%%%%%%%%%%%%%%%%%%%%%%%%%%%%%%%%%%%%%%%%%%%%%%%%%%%%%%%%%%%%%%

In Fig. \ref{Fig:figure5} the difference between the LSDA and DMFT total Compton profiles 
$\Delta J$ is shown together with its second order moment $p^2 \Delta J$. 
The theoretical CP profiles were broadened with a Gaussian of 0.22 a.u. width. 
The momentum dependence of $p^2 \Delta J(p)$ and the second 
moments of the difference Compton profiles are depicted in Fig.\ \ref{Fig:figure5}
along [001] and in Fig.\ \ref{Fig:figure5} for the [110] and [111] directions. 
For all principal directions the weight of $p^2 J^{DMFT}(p)>p^2 J^{LSDA}(p)$ for moments 
$p_z < 2$ a.u., while for larger moments $p^2 \Delta J(p)$ has a negative weight. This 
shows that the DMFT derived correlation is significant in the low momentum region, 
$p_z < 2$ a.u.. This result allows to extend the concept of spectral weight transfer 
from real space into momentum space. In the real space representation spectral weight 
transfer is discussed in terms of the changes of the spectral properties as a 
function of the strength of the local interaction, $U$. For larger $U$ values, 
spectral weight is redistributed form the Fermi level towards high binding energies.
In momentum space
more states/electrons are transferred towards low energies for small momenta $p_z < 2$ a.u.,
while at large values of $p$ spectral weight is shifted towards high momentum region. 

%%%%%%%%%%%%%%%%%%%%%%%%%%%%%%%%%%%%%%%%%%%%%%%%%%%%%%%%%%%%%%%%
\begin{figure}[h]
%\vspace{0.5cm}
%\begin{center}
%\hspace*{0.2cm}
   \includegraphics[width=0.99\linewidth, clip=true]{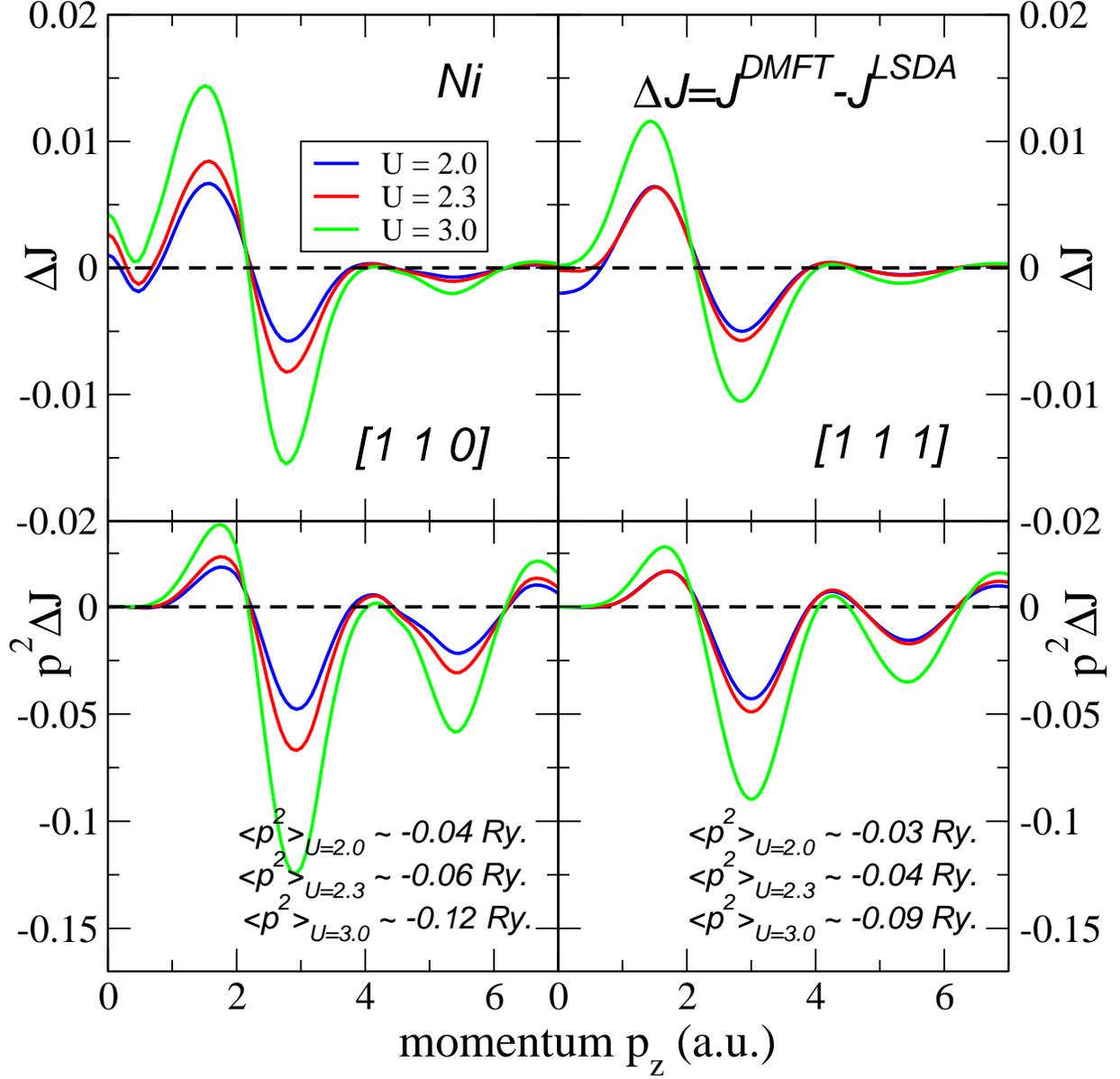}
%\end{center}
\caption {\label{Fig:figure6}(Color on-line) Differences between the LSDA and LSDA+DMFT total 
Compton profiles of Ni along the [110] (left column) and [111] (right column) directions
for different values of $U$. The values of the second moment are larger along the [110]
direction at all values of $U$.
}
\end{figure}
%%%%%%%%%%%%%%%%%%%%%%%%%%%%%%%%%%%%%%%%%%%%%%%%%%%%%%%%%%%%%%%%

Along the [001] direction the spectra have mostly a negative weight, and
the estimated values for the second moments are -0.05 Ry for $U$ = 2 eV, 
-0.07 Ry for $U$ = 2.3 eV and -0.15 Ry for $U$ = 3.0 eV. One notices that larger values 
of $U$ determine larger values (in absolute terms) of the second moment, although the
precise increase of these values is different along different directions. Along the [110] direction
the magnitude of the second moment is similar being -0.04, -0.06, -0.12 Ry, while along
[111] smaller values for the second moments are obtained. 
%For Ni the strongest second moment weight is seen along the [110] direction.

\subsection{Discussions and Conclusion}
\label{second+E}
The Compton scattering experiment yields the one-dimensional momentum 
distribution for the scatterer. Therefore, it is possible to use Compton data
to calculate the expectation values of operators which are functions of momentum
$\langle p^n \rangle$. The value for $n=2$ is of special interest, since 
$1/2 \langle p^n \rangle$ gives the electronic kinetic energy,
leading to a connection with the total energy of the scattering system. 
As a result, the computed Compton profile can be easily interpreted as a
very fundamental quantity. In the following we discuss the connection between 
the second moment of the difference between correlated (LSDA+DMFT) and non-correlated
(LSDA) Compton profiles and the kinetic energy of the electronic system. Our main focus
is on the bond average of the second moment of the difference Compton profiles: 
\begin{equation}
\overline{\langle p^2 \rangle}=\frac{1}{N_b} \sum_{\bf K} \int_0^{\infty} p_z^2 
\Delta J_{\bf K}(p_z) dp_z \propto E_{kin}^{DMFT} - E_{kin}^{LSDA} .
\label{eq:kkr_lmto}
\end{equation}
Here the overbar represents the average taken over the bonds extended along the {\bf K}-directions, 
$\Delta J_{\bf K}(p_z)$ is the difference of total Compton profile, $N_b$ is the number of bonds
and the energies on the right hand side are the kinetic energies computed in DMFT/LSDA.
In general, calculating total energies in LSDA+DMFT is a difficult task, and requires
the evaluation of an energy functional with several terms~\cite{KS06,MM09}
including the Galitskii-Migdal contribution~\cite{GM58}, i.e. $1/2 \text{Tr}[\hat{\Sigma} \hat{G}]$,
and the double counting as well.
The LSDA+DMFT total energy functional can in principle be analyzed to obtain an expression
for the kinetic energy similarly to what is done for DFT~\cite{VIR85,VIR09}. When focusing
on the differences between LSDA+DMFT and LSDA, one can write:
\begin{align}
\Delta E_{kin} = & \text{Tr}\left[{\hat{H}_{{KS}}^{DMFT}\hat{G}^{DMFT}}\right] - \nonumber \\
		 & \text{Tr}\left[{\hat{H}_{{KS}}^{LSDA}\hat{G}^{LSDA}}\right] +
		  \langle \Delta V_{KS} \rangle + \langle \Delta T_c \rangle. 
\label{eq:kinetic_en}
\end{align}
In this expression the first and second terms on the right hand side are the single particle
energies of the Kohn-Sham Hamiltonian within LSDA+DMFT and LSDA, while the third term is
the expectation value of the difference of their corresponding Kohn-Sham potentials. 
The last term in Eq.\ (\ref{eq:kinetic_en}) is the variation of the exchange-correlation
contribution to the kinetic energy, and can in principle be expressed in terms of the
exchange-correlation potential and its gradient~\cite{VIR09}.  

In spite of the recent progress in improving the accuracy of LDA+DMFT energetics ~\cite{KS06,MM09}
it is still a difficult task to compute not only LSDA+DMFT total energies but also the terms 
in discussion with a high degree of accuracy. In addition, the energy components given in 
Eq.\ (\ref{eq:kinetic_en}) being complete traces, would not provide any information about the 
magnitude of the correlation energy along different bonds or directions in the ${\bf k}$-space. 
On contrary, the analysis of the second moments along the bonds $\langle p^2 \rangle_{\bf K}$  
as shown in Figs.\ \ref{Fig:figure2}, \ref{Fig:figure5} and \ref{Fig:figure6}
demonstrates that changes in the kinetic energy because of electronic correlations are anisotropic. 
One has to note that the main source for the anisotropy in the momentum space is bond
directionality that is already captured within the LSDA. However this can not provide
any measure of the electronic correlations.
 
%%%%%%%%%%%%%%%%%%%%%%%%%%%%%%%%%%%%%%%%%%%%%%%%%%%%%%%%%%%%%%%%%
\begin{figure}[h]
%\vspace{0.5cm}
%\begin{center}
%\hspace*{0.2cm}
%  \includegraphics[width=0.99\linewidth, clip=true]{Comparison_p2J_totalE_Ni_v3.eps}
  \includegraphics[width=0.99\linewidth, clip=true]{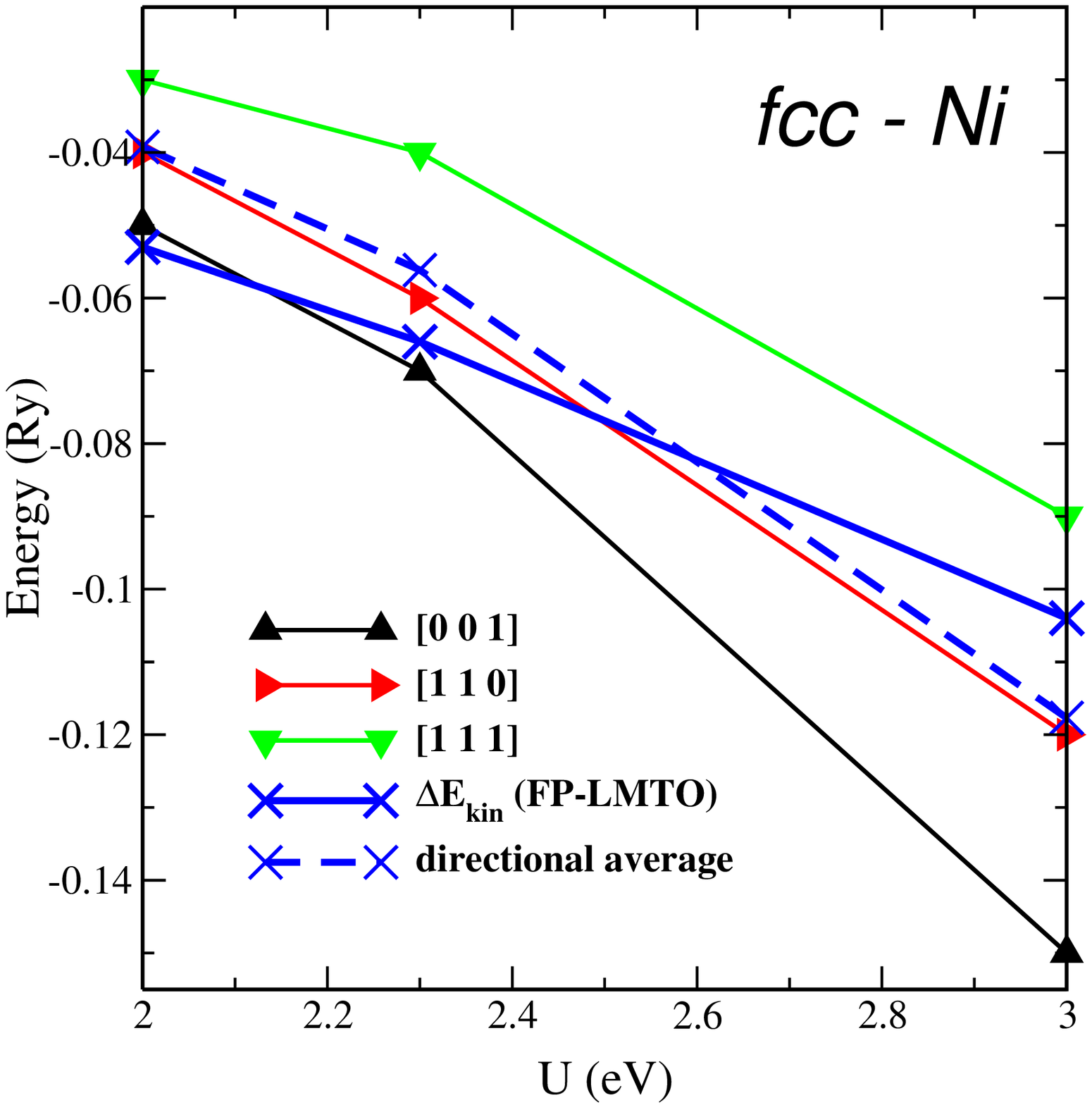}
%\end{center}
\caption {\label{Fig:figure7}(Color on-line) Second moment of the difference
$\Delta J_{\bf K}(p)$ along the directions [001] (solid black line with triangles up),
[110] (solid red line with triangles right), and [111] (solid green line with triangles down),
together with the directional average over the bonds (dashed blue line with crosses). Moreover
the approximate difference in kinetic energy between LSDA+DMFT and LSDA as obtained by
FP-LMTO~\cite{GM07,GM12} is also shown. 
}
\end{figure}
%%%%%%%%%%%%%%%%%%%%%%%%%%%%%%%%%%%%%%%%%%%%%%%%%%%%%%%%%%%%%%%%%
 
In Fig.~\ref{Fig:figure7} we show the second  moment of the difference $\Delta J_{\bf K}(p)$
along the principal directions and its directional average. The latter is estimated as the 
weighted sum of the nearest neighbors i.e.\ six times the contribution along [001], 12 times
the contribution along [110] and 8 times the contribution along [111] divided by the total
number of neighbors (26). The values of the second moments are almost similar along [110]
and [001] directions and smaller than along [111] direction. 
For sake of comparison, in Fig.~\ref{Fig:figure7} we also show the 
variation of the kinetic energy as obtained from Eq.~(\ref{eq:kinetic_en}) by
ignoring the last term $\langle \Delta T_c \rangle$. These data were obtained
through a full-potential linearized muffin-tin orbital (FP-LMTO) code~\cite{GM07,GM12},
which has been shown to give results in very good agreement with SPR-KKR regarding 
LSDA+DMFT total energies~\cite{MM09}.
Due to the aforementioned approximations and a very high sensitivity to the numerical 
and computational details, any quantitative comparison between second moment of $\Delta J_{\bf K}(p)$ 
from SPR-KKR and $\Delta E_{kin}$ from FP-LMTO is problematic. In fact, from 
Fig.~\ref{Fig:figure7}, it is clear that the two contributions are still far from a
quantitative agreement. However, we capture a consistent qualitative picture pointing 
to a decrease of kinetic energy difference for increasing U. It is important to note
that the second moment of the Compton profile turns out to be negative because the 
positive contribution in the low momentum region up to $p_z < 2$ a.u., is completely 
overruled by the large negative contribution from high momenta. In the low momentum 
region increasing the values of $U$ an increasing in the kinetic energy is obtained, which 
is in agreement with the argument that the presence of $U$ penalizes the electrons and 
leads to an increase in their kinetic energy. This argument is not valid any more in the 
region of high momenta, where the mean-field type exchange-correlation dominates 
the ``Hubbard-U'' contribution. Further analysis is needed in order to make a more 
quantitative comparison, especially to understand the role of the double counting 
correction and the effects of the expectation value of $T_c$, discarded in the 
present analysis. 

Concerning the charge redistribution in Ni, the real space picture corresponds to 
the contraction of the electronic charge because of correlation effects, as seen 
from previous coordinate space charge computation~\cite{CVA+03}. The overall negative
second moment of the difference tells us that the corresponding kinetic energy is 
decreasing with increasing the strength of U. Therefore, the less mobile correlated
d-electrons would weaken the covalent component of the metallic bonding seen 
in transition metals. In addition as one can see in Fig. \ref{Fig:figure7}, this effect
is anisotropic. 
Similarly for Fe the covalent d-d contribution is weakened as well in the presence of 
correlations. In comparison to Fe, Ni shows 
larger values of the second moments $p^2 \Delta J$ along [001] and [110], while for 
the [111] direction the opposite situation is found. This is an expected result as 
the shortest distance and the strongest bond is realized along the [111] direction, 
of the bcc-structure. As a common feature both Fe and Ni show positive values
for $\Delta J_{\bf K}(p)$ and consequently positive second order moments  
up to the $p_z < 2$ a.u. momenta. This value seems to be the upper bound in the 
momentum space up to which many-body correlation effects captured by DMFT have a 
larger weight in comparison to the exchange and correlation effects described by 
LSDA. 

%\section{Conclusions}
To conclude, in the present work we studied the influence of electronic correlations 
on the Compton profiles of Fe and Ni within the framework of density functional theory 
using  the LSDA+DMFT approach. 
The second moment of the Compton profile difference allows to quantify
the momentum space anisotropy of the electronic correlations of Fe and Ni.
The changes in the shape and magnitude of the anisotropy have been discussed
as a function of the strength of the Coulomb interaction $U$. According to our
results Ni has a larger momentum space anisotropy of the second moment of
the total Compton profile in comparison with Fe. In the range of the studied
values of $U$ no significant dependence is seen in the anisotropy of the Compton
profile. As an overall conclusion DMFT introduces moderate improvements for the
spectral features in particular at low momentum. Further progress is needed to 
bridge between momentum density and the total energy of the system through the 
computed Compton profile.

\section{acknowledgments}
This research was supported in part by the Deutsche Forschungsgemeinschaft 
through FOR 1346. DAAD and the CNCS - UEFISCDI (project number PN-II-ID-PCE-2012-4-0470) 
are gratefully acknowledged. D. B would like to express a deep gratitude 
to Prof. V. Cri\c san (UBB Cluj, Romania) for his support and encouragement 
during her study period and early career.
I. D. M acknowledges the Swedish National Infrastructure 
for Computing (SNIC) for providing computational resources at the National Supercomputer 
Center (NSC). L. C. acknowledge the discussions with I. Leonov.
    
\bibliography{paper}

\end{document}